\documentstyle[12pt]{article}
\begin{document}
\newcommand{\preprint}[1]{\begin{table}[t] 
           \begin{flushright}              
           \begin{large}{#1}\end{large}    
           \end{flushright}                
           \end{table}}

\baselineskip 18pt
\preprint{TAUP-2166-94}
\title{Coordinate System In Quantum Gravity}
\author{N.Itzhaki\\Raymond and Beverly Sackler Faculty of  Exact Sciences
\\School
 of Physics and Astronomy\\Tel Aviv University, Ramat Aviv, 69978, Israel}
\maketitle
\begin{abstract} We discuss a Gedanken experiment in which we construct a
 physical coordinate system which covers the universe.
Using general properties of
 quantum gravity we find that the minimum uncertainty of the coordinate system
 is $\sqrt[5]{R}$ where $R$ is the radius of the universe.\end{abstract}
\newpage

The existence of a minimum length at Planck's scale has been proposed from
the study of thought experiments  which are model independent
  \cite{ma,me,de,ah} or through an analysis of collisions at planckian energies
in the context of string theories\cite{ve,ko}
(for recent review see  \cite{ga}).
In \cite{it} we  studied the problem of time measurement in quantum gravity.
We have shown  that one can
not synchronize two clocks to an accuracy better than
 \(\Delta T=\sqrt{8\log\frac{x}{x_{c}}}\)
(in units where $\hbar =c=G=1$), where $x_{c}$ is the shortest distance for
 which general relativity
is a good approximation to quantum gravity, and $x$ is
 the distance between the clocks.
The fact that $\Delta T$ is not just a constant of nature of the order of the
Planck scale as one might expect,
 but an increasing unbounded function of $x$ indicates
 that the problem of finding a quantum theory of gravity is more intricate than
 the problem of renormalization at the Planck scale.
The reason that one might not be too concerned about it, is that the distance
 between the clocks is bounded by the diameter of the universe ($\approx
 10^{60}$).
Therefore, even if one assumes that $x_{c}=1$ one  gets
 $\Delta T_{max}\approx 33$, which is not too far from $1$, yet for an open
 universe $\Delta T_{max}=\infty $.
In this letter we would like to generalize the two-clock problem mentioned
 above by studying a physical coordinate system (p.c.s.) \footnote{By p.c.s.
we mean any physical device which can measure any event or events
which occur at the universe with some accuracy.} that covers the universe.
In particular we would like to investigate the
dependence of the accuracy of the p.c.s. on the size of the universe.

While in general relativity the clocks and the rods,
which are used to construct
 a p.c.s., must  have an infinitesimal energy momentum tensor in
 order not to change the metric,
in quantum theories without gravitation we must
 use p.c.s. with $\Delta P_{\mu}\rightarrow \infty $ in order to get
 $\Delta X_{\mu}\rightarrow 0 $.
Although theoretically, there is no limitation on the accuracy of the p.c.s.
  in quantum theories without gravitation, in practice the energy in the
 universe is finite.
Thus there is a practical limitation on the accuracy of a p.c.s. in
 quantum theories even when gravitation is absent.
In other words--- in classical theories we can use infinitesimal energy of the
 universe in order to construct coordinate system for the whole universe with
 infinitesimal uncertainty,
while in quantum theories even if we use the whole energy in the universe to
 construct
a coordinate system, we still have, due to the uncertainty principle,
 only a finite  accuracy.

In quantum gravity the problem in constructing a p.c.s. is much more
 intricate, since, as we shall see later on, even theoretically the accuracy of
 the p.c.s. is bounded from below.

In order to build a p.c.s. with accuracy $b$, \footnote{Since we assume the
weak field approximation $b$ is noting but  the bare accuracy of the rods
and clocks which are used to construct the p.c.s.}  which covers
 the whole universe, one must construct a device which is able to measure the
location at which any events in the universe occur with accuracy $b$
therefore  one must  divide the universe into 4-dimensional cells with
  volume $b^{4}$.
When
one considers quantum theory one finds that due to the uncertainty principle
 $\Delta P_{\mu}\geq
 b^{-1}$  in each cell.
In the weak field approximation \begin{equation}
 g_{\mu\nu}=\eta_{\mu\nu}+h_{\mu\nu} \;,\;\mid h_{\mu\nu}\mid\ll 1
 \end{equation}
we have
\begin{equation} P_{\mu}=\int d^{3}xT_{0\mu}\end{equation}
where $T_{\mu\nu}$ is the energy momentum tensor. Thus \begin{equation}\Delta
 \int_{cell} d^{3}xT_{0\mu}>b^{-1}\end{equation}
in each cell.
Note that eq.(3) is not a consequence of vacuum fluctuation in each cell, but a
 constraint
that the energy momentum tensor of the physical clocks and rods must
 satisfy in order to have accuracy $b$.

There is a well know solution to the field equation in the weak field
 approximation where one works  in an harmonic coordinate system (for instance
\cite{we}) \begin{equation}g_{\mu\nu}(x,t)=\eta_{\mu\nu}+4\int
 d^{3}x^{'}\frac{S_{\mu\nu}(x^{'}, t-\mid x-x^{'}\mid )}{\mid x-x^{'}\mid
 }\end{equation}
where
 \begin{equation}S_{\mu\nu}=
T_{\mu\nu}-\frac{1}{2}\eta_{\mu\nu}T^{\gamma}_{\gamma}.\end{equation}
Since the uncertainties of any two cells are independent of each other we get
\begin{equation}\Delta^{2}g_{\mu\nu}(x,t)=16\sum_{cells}\frac{(\Delta\int
 d^{3}x_{'} S_{\mu\nu})^{2}}{(x-x_{cell})^{2}}\end{equation}
 $S_{0i}=T_{0i}$ ;  thus  we have in each cell\begin{equation}\Delta\int
 d^{3}xS_{0i}\geq b^{-1}.\end{equation}
Note that $\Delta N_{cell}\approx \frac{dV}{b^{3}}$.
Since  $b\ll R$ ,   we can replace the sum in eq.(6) by an integral, and use
 eq.(7) to obtain \begin{equation} \Delta^{2}g_{0i}=16\int_{0}^{\pi}
 d\theta\int_{0}^{2\pi} d\varphi\int_{0}^{R}
 dr\frac{\sin\theta}{b^{5}}\end{equation}
therefore\begin{equation}\Delta g_{0i}\geq
 \frac{8\sqrt{\pi}}{b^{2}}\sqrt{\frac{R}{b}}.\end{equation}
The invariant distance is $ds=\sqrt{g_{\mu\nu}dx^{\mu}dx^{\nu}}$ ,
hence the real accuracy of the p.c.s. is not $b$ but ,
 \begin{equation}max
\{\Delta (\sqrt{g_{\mu\nu}dx^{\mu}dx^{\nu}})\}\end{equation}
Denoting the p.c.s. accuracy by $d$ one has :\begin{equation} d\geq \max
 (b,\sqrt[4]{\frac{R}{b}})\geq \sqrt[5]{R}.\end{equation}
Note that although we calculate $d$ in a particular choice of coordinate $d$
as defined in eq.(10) is invariante under coordinate transformation,
hence the conclusions are independent of the choice of coordinates.
The reason why we did not calculate the exact numerical constant in eq.(10) is
 that the entire calculation was based on the weak field approximation, while
 for $b=\sqrt[5]{R}$ we get from eq.(9) $\Delta g_{0i}>1$.
Thus one can consider the final result only as a qualitative result
 \begin{equation}d\propto \sqrt[5]{R}.\end{equation}

The meaning of $\sqrt[5]{R}$ as a lower bound to the accuracy of the p.c.s.
 is the following :
 if one would try to construct a p.c.s. using clocks and rods which
 are more accurate then  $\sqrt[5]{R}$, than  the quantum fluctuation  of their
 accuracy would be larger then the  their bare accuracy , and the classical
 description of the p.c.s. will be violated.
Notice that  for $R=10^{60}$ ,   the energy density of the p.c.s. is
 $R^{\frac{-4}{5}}\approx 10^{-48} $ at the minimum accuracy.
While the energy density of the universe is only $\approx 10^{-123}$, thus the
 classical description of the large scale universe is valid.
Since the weak field approximation is not valid in the early universe a further
 investigation is needed
 in order to find out, when in the early universe the quantum corrections play
 an essential role
and the classic description of cosmology is no longer valid.
Still,
it is tempting to use the above calculation , which was based on the weak
 field approximation
in order to compare between the accuracy $d$ and the wave length of the
 background radiation $l_{bac}\approx \frac{R}{\sqrt[3]{N}}$;
we find that $l_{bac}=d$ at $R\approx N^{\frac{5}{12}}\approx 10^{33}$ which is
 approximately the size of the universe at the big bang.

Note that $\sqrt[5]{R}$ is the local accuracy of the p.c.s. i.e., the
 minimum invariant distance of each cell.
Yet it is not the uncertainty in the invariant distance between any two events
 in the universe,
which is the global accuracy of the p.c.s..
In order to calculate the global accuracy one needs to remember that when a
 coordinate
system is constructed of cells of length $d\pm \Delta d$, the uncertainty
 in the distance between two events which are separated by $N$ cells is
 $d+\sqrt{N}\Delta d$.
In our case $d=b$ , $N_{max}=\frac{R}{b}$ and $\Delta d=\Delta g_{0i}b$.
Therefore we find that the global accuracy of the p.c.s. is
\begin{equation} d_{gl}=b+8\sqrt{\pi}\frac{R}{b^{2}}\end{equation}
The minimum is \begin{equation} d_{gl,min}\approx \sqrt[3]{R}\end{equation}
Note that at the minimum $\Delta g\approx \frac{1}{\sqrt[3]{R}}$,
since $R\gg 1$
 the weak field approximation is valid.

While the fact that the global accuracy is a function of a global variable $R$
 is not surprising,
 the fact that
 the
local accuracy in quantum gravity  is, operationally a function of a global
 variable is surprising and somewhat disturbing,
since by definition, coordinate system is a local concept.
In
our opinion this result gives rise to the question whether it is possible  to
 describe quantum gravity by means of local quantum field theory.

\vspace{1cm}
I would like
to thank Prof. Y.Aharonov and Prof. S.Nussinov for useful comments,
 and especially Prof. A.Casher for helpful discussions.


\newpage
\begin{thebibliography}{99}
\bibitem{ma} M.Maggiore Phys.Lett.B 304 (1993) 65.
\bibitem{me} C.A.Mead Phys.Rev.135 B849 (1964).
\bibitem{ve} D.Amati, M.Ciafaloni and G.veneziano, Phsy. Lett. B216 (1989) 41.
\bibitem{ah} D.V.Ahluwalia gr-qc/9308007.
\bibitem{de}
B.S.Dewitt, in Gravitation, ed. L.Witten (John Wiley \&sons New-york).
\bibitem{un}
W.G.Unruh, in Quantum Theory of Gravity, ed. S.Christensen (Adam Hilger).
\bibitem{ga} L.J.Garay gr-qc/9403008, Imperial/TP/93-94/20.
\bibitem{ko} K.Konishi, G.Paffuti and P.Provero, Phsy. Lett. B234 (1990) 276.
\bibitem{it} N.Itzhaki  Phys.Lett.B 328 (1994) 274.
\bibitem{we}
S.Weinberg Gravitation and Cosmology (John Wiley \& sons New-york  1972).
\end{thebibliography}
\end{document}